\documentclass[aps,prb,groupedaddress,showpacs,showkeys,preprint]{revtex4}

\usepackage{natbib}
\usepackage{graphicx}
\usepackage{amsmath}
\begin{document}
\title{Universal chiral-triggered magnetization switching in confined nanodots.}

\author{Eduardo~Martinez$^1$  \footnote{Corresponding author's e-mail: edumartinez@usal.es}, Luis~Torres$^1$, Noel~Perez$^1$, Maria~Auxiliadora Hernandez$^1$, Victor~Raposo$^1$ and Simone~Moretti$^1$} 

\affiliation{$^1$ Universidad de Salamanca. Plaza de los Caidos s/n, E-37008, Salamanca. Spain.}

\date{\today}

\begin{abstract}

\textbf{Spin orbit interactions are rapidly emerging as the key for enabling efficient current-controlled spintronic devices. Much work has focused on the role of spin-orbit coupling at heavy metal/ferromagnet interfaces in generating current-induced spin-orbit torques. However, the strong influence of the spin-orbit-derived Dzyaloshinskii-Moriya interaction (DMI) on spin textures in these materials is now becoming apparent. Recent reports suggest DMI-stabilized homochiral domain walls (DWs) can be driven with high efficiency by spin torque from the spin Hall effect. However, the influence of the DMI on the current-induced magnetization switching has not been explored nor is yet well-understood, due in part to the difficulty of disentangling spin torques and spin textures in nano-sized confined samples. Here we study the magnetization reversal of perpendicular magnetized ultrathin dots, and show that the switching mechanism is strongly influenced by the DMI, which promotes a universal chiral non-uniform reversal, even for small samples at the nanoscale. We show that ultrafast current-induced and field-induced magnetization switching consists on local magnetization reversal with domain wall nucleation followed by its propagation along the sample. These findings, not seen in conventional materials, provide essential insights for understanding and exploiting chiral magnetism for emerging spintronics applications.}

\end{abstract}


\maketitle

Understanding and controlling the current-induced magnetization dynamics in high perpendicular magnetocristaline anisotropy heterostructures consisting of a heavy-metal (HM), a ferromagnet (FM) and an oxide (HM/FM/O) or asymmetric HM1/FM/HM2 stacks, is nowadays the focus of active research\cite{Miron_11a,Liu_11,Liu_12a,Liu_12b,Garello_13,Garello_14,Miron_11b,Haazen_13,Emori_13,Ryu_13,Emori_13b,Je_13,Ryu_14,Hrabec_14,Franken_14,Torrejon_14}. Apart from their interest for promising spintronics applications, these systems are also attracting growing attention from a fundamental point of view due to the rich physics involved in the current-induced magnetization switching (CIMS)\cite{Miron_11a,Liu_11,Liu_12a,Liu_12b,Garello_13} and in the current-induced domain wall motion (CIDWM)\cite{Miron_11b,Haazen_13,Emori_13,Ryu_13,Emori_13b}. Indeed, the combination of a HM and a thin FM film gives rise to new phenomena which normally vanish in bulk, but play an important role as the thickness of the FM is reduced to atomistic size.

Current-induced torques arising from spin-orbit phenomena can efficiently manipulate magnetization. In particular, the Slonczewski-like spin-orbit torque (SL-SOT)\cite{Miron_11a,Liu_11,Liu_12a,Liu_12b,Garello_13,Garello_14,Miron_11b,Haazen_13,Emori_13,Ryu_13, Emori_13b} can switch the magnetization from up ($\uparrow$) to down ($\downarrow$) states and vice versa under the presence of small in-plane fields. The SL-SOT is expressed as

\begin{equation}
\vec{\tau}_{SL}=-\gamma_{0}H_{SL}\vec{m}\times (\vec{m}\times \vec{\sigma}) \label{eq:SLSOT}
\end{equation}

\noindent where $\gamma_{0}$ is the gyromagnetic ratio, $\vec{m}$ the unit vector along the magnetization, $\vec{\sigma}=\vec{u}_{z}\times \vec{u}_{j}=\vec{u}_{y}$ the unit vector along the polarized current which is perpendicular to both the easy axis ($z$) and current direction given by $\vec{u}_{j}$, and $H_{SL}$ parameterizes the torque. CIMS in ultrathin Pt/Co/AlO, where the Co layer is only $0.6\mathrm{nm}$ thick (around three atomic layers), was experimentally observed first by Miron and coworkers\cite{Miron_11a}, where the switching was attributed to SL-SOT due to the Rashba field\cite{Yu_84,Manchon_08}. The Rashba effect would generate both field-like (FL-SOT)\cite{Yu_84,Manchon_08} and Slonczewski-like (SL-SOT)\cite{Wang_12,Kim_12} spin-orbit torques. Similar to the conventional spin transfer torque (STT)\cite{Thiaville_05}, both Rashba FL and SL SOTs have magnitudes proportional to the spin polarization of the current ($P$) flowing through the FM, and therefore, they are expected to be negligible for an ultrathin FM, as reported in experimental studies\cite{Cormier_10,Emori_12,Tanigawa_13}. Indeed, Liu et al.\cite{Liu_12a} studied CIMS in Pt/Co/AlO, similar to the study by Miron et al.\cite{Miron_11a} but they did not find any significant dominant Rashba FL torque, and therefore the Rashba contribution to the SL-SOT should be even vanishingly small. This was also the conclusion from switching experiments in asymmetric Pt/Co/Pt\cite{Haazen_13} and for Pt/CoFe/MgO\cite{Emori_13}. Instead of the Rashba SL-SOT, the switching is consistent with an alternative SL-SOT based on the spin Hall effect (SHE)\cite{Dyakonov_71,Hirsch_99}. The SL-SOT due to the SHE is physically distinct from other torques STTs and Rashba-SOTs: it is independent of $P$ because it arises from the spin current generated in the HM, rather than the spin polarization of the charge current in the FM.

The key to the existence of the SOTs is a high spin-orbit coupling combined with structural inversion asymmetry (SIA) in these heterostructures: if the top and bottom interfaces/layers sandwiching the FM were completely symmetric, all the mentioned effects should cancel out. However, not only the SIA plays a role in these current-induced magnetization dynamics but, it can also influence the static magnetization state through the interfacial Dzyaloshinskii-Moriya interaction (DMI)\cite{Moriya_60,Bode_07,Heide_08,Thiaville_12}. In systems with SIA, the interfacial DMI is an anisotropic exchange contribution which directly competes with the exchange interaction, and when strong enough, it promotes non-uniform magnetization textures of a definite chirality such as spin helixes\cite{Chen_13}, chiral domain walls (DWs)\cite{Thiaville_12,Haazen_13,Emori_13,Ryu_13,Emori_13b} and skyrmions\cite{Sampaio_13,Tomasello_14,Rohart_13}. In particular, the experiments on current-induced DW motion along Pt/Co/AlO\cite{Miron_11b} or Pt/CoFe/MgO\cite{Emori_13,Emori_13b} can be explained by the combined action of the DMI and the SHE. The strong DMI in these Pt systems is the responsible of the formation of the Neel walls with a given chirality, which are driven by the SHE\cite{Emori_13,Ryu_13,Emori_13b}. However, the influence of the DMI on the CIMS has not been explored nor is yet well-understood, due in part to the difficulty of disentangling spin torques and spin textures in nano-size confined dots.

On the other hand, experiments on CIMS in these asymmetric multilayers are usually interpreted in the framework of the single-domain model (SDM) which neglects both the exchange and DMI contributions, and only a few recent studies in extended samples at the microscale ($15\mathrm{\mu m} \times 1.2\mathrm{\mu m}$) have considered the non-uniform magnetization by full 3D micromagnetic simulations\cite{Finocchio_13,Lee_14,Perez_14,Pizzini_14}. Here we focus on CIMS of a ultrathin Pt/Co/AlO with in-plane dimensions two orders of magnitude below ($\approx 100\mathrm{nm}$). Although these dimensions should be amenable for the uniform magnetization description, our study indicates that the DMI is also essential to describe the CIMS at these dimensions, which occurs through chiral asymmetric DW nucleation and propagation. We analyze the key ingredients of the switching and confirm that a full micromagnetic analysis is necessary to describe and quantify the spin Hall angle under realistic conditions.

\textbf{RESULTS}

\begin{figure}
\includegraphics[width=0.80\textwidth]{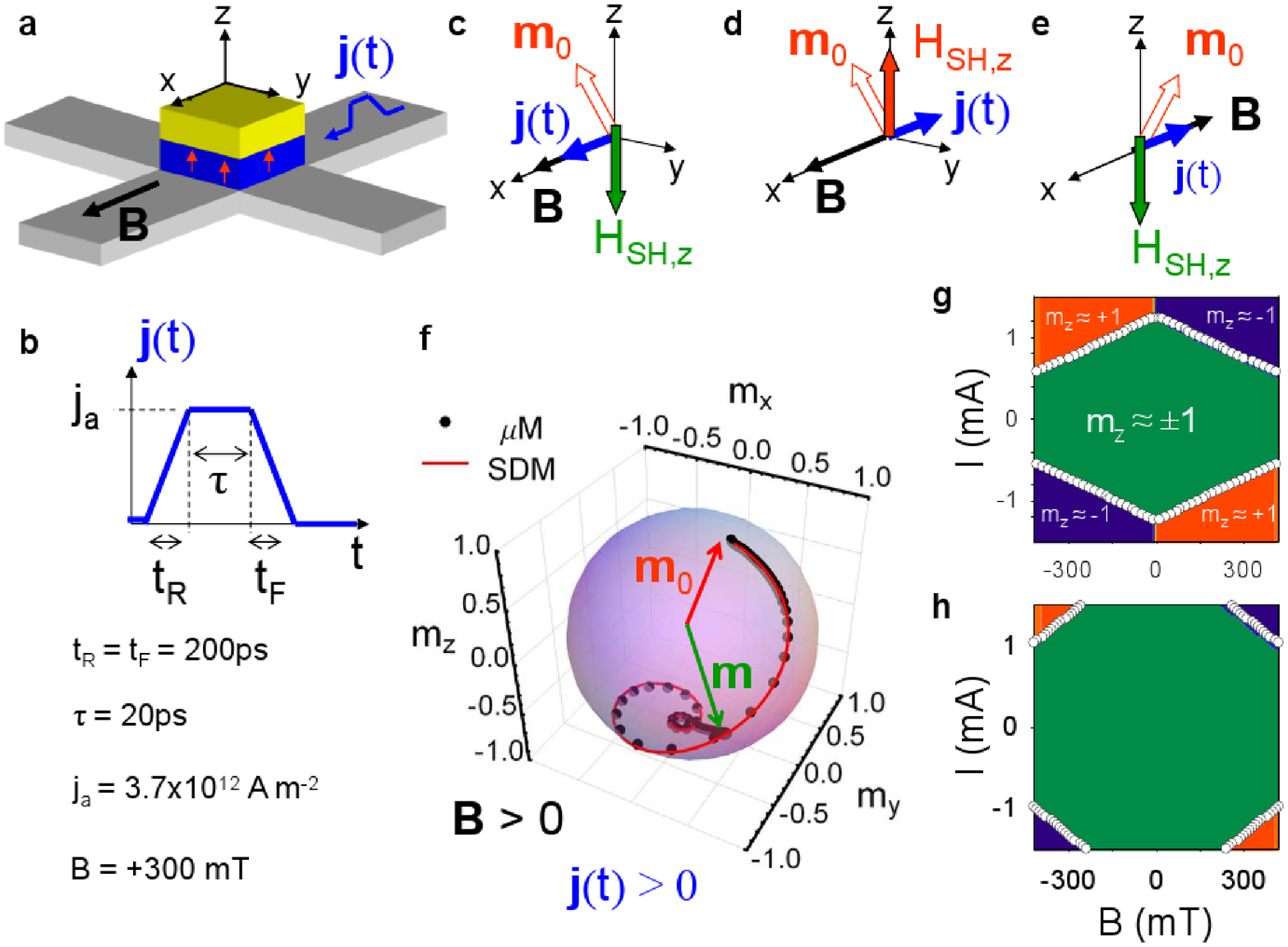}
\caption{\textbf{Current induced magnetization switching in the absence of DMI ($D=0$)} $|$ (a) Schematic representation of the analyzed heterostructure with the Co layer in blue. (b) Temporal variation of the density current pulses $j(t)$ with rising ($t_{R}$), falling ($t_{F}$) and duration ($\tau$) times. (c)-(e) Out-of-plane component of spin Hall effective field $H_{SH,z}$ as a function of the applied field $B$ and density current $j(t)$ directions. (f) Magnetization trajectories starting from up state ($\uparrow$) to down state ($\downarrow$) under a static field of $B=300\mathrm{mT}$ and a pulse with $t_{R}=t_{F}=200\mathrm{ps}$, $\tau=20\mathrm{ps}$ and $I=1.2\mathrm{mA}$ ($j_{a}=3.7\times 10^{12}\mathrm{A/m^{2}}$ flowing through both the Pt and Co layers) for $\theta_{SH}=0.2$. Solid red line depicts Single Domain Model (SDM) results whereas solid black dots correspond to full micromagnetic ($\mu M$) simulations in the absence of DMI (D=0) for the averaged magnetization components over the sample volume ($<...>$). (g)-(h) Stability phase diagrams indicating the terminal out-of-plane magnetization component $m_{z}$ as a function of $B$ and $j(t)$ for $t_{R}=t_{F}=200\mathrm{ps}$, $\tau=20\mathrm{ps}$, and $T=300\mathrm{K}$ as computed with the SDM with a high $\theta_{SH}=0.2$ (g) and with realistic $\theta_{SH}=0.11$ (h). Open circles denote the transition between switching and no-switching at zero temperature.}
\label{fig:Fig1}
\end{figure}

The considered heterostructure here consists on a thin ferromagnetic Co nanosquare with a side of $L = 90\mathrm{nm}$ and a thickness of $L_{z}=0.6\mathrm{nm}$ sandwiched between a AlO layer and on top of a Pt cross Hall (Fig. \ref{fig:Fig1}(a)). The thickness of the Pt layer is $3\mathrm{nm}$. Typical high PMA material parameters were adopted in agreement with experimental values\cite{Pizzini_14,Garello_13,Garello_14}. Details about the physical parameters can be found in Methods.

\textbf{Cuasi-uniform current-induced magnetization switching in the absence of the DMI: single domain approach and micromagnetic results.} The current induced magnetization dynamics under static in-plane longitudinal field $\vec{B}=B\vec{u}_{x}$ and current pulses $\vec{j}(t)=j(t)\vec{u}_{x}$ is studied from both Single Domain Model (SDM) and full micromagnetic Model ($\mu M$) points of view (see Methods). We first review the CIMS in the framework of the SDM, where the magnetization is assumed to be spatially uniform ($\vec{m}(t)=(m_{x}(t),m_{y}(t),m_{z}(t))$). Within this approach the conventional symmetric exchange and interfacial DMI are not taken into account ($D=0$). In the absence of in-plane fields ($B=0$) or thermal fluctuations, with the magnetization initially pointing along the easy $z$-axis ($\vec{m}(0)=\pm\vec{u}_{z}$, $\uparrow / \downarrow$), a moderate current $j$ along the longitudinal direction ($x$-axis) only generates an effective SHE field along the $x$-axis which does not promote the out-of-plane magnetization reversal ($\vec{H}_{SH}=H_{SH}(j)\vec{m}\times\vec{\sigma}=-H_{SH}(j)m_{z}\vec{u}_{x}$). However, in the presence of a longitudinal field $\vec{B}=B\vec{u}_{x}$ below the saturating in-plane field ($B_{sat}\approx 1\mathrm{T}$), $\vec{m}$ acquires a finite longitudinal component $m_{x}\neq 0$ parallel to $\vec{B}$, and the current pulse $j(t)$ generates an out-of-plane component effective SHE field $H_{SH,z}=H_{SH}(j)m_{x}$. If $\vec{j}(t)$ is parallel to $\vec{B}$ (either $j(t)>0$ and $B>0$ as in Fig. \ref{fig:Fig1}(c), or $j(t)<0$ and $B<0$ as in Fig. \ref{fig:Fig1}(e)), and their magnitudes are sufficiently strong, the magnetization is stabilized pointing parallel to the out-of-plane component of $H_{SH,z}=H_{SH}m_{x}<0$, i.e. along the $-z$-axis (Fig. \ref{fig:Fig1}(c) and (e)). On the contrary, if the field and the current pulse are anti parallel to each other (either $j(t)>0$ and $B<0$, or $j(t)>0$ and $B>0$), $\vec{m}$ is stabilized along the $+z$-axis (Fig. \ref{fig:Fig1}(d)).

Fig. \ref{fig:Fig1}(f) shows the 3D magnetization trajectories for CIMS starting from the up state ($\uparrow$) with $j(t)>0$ and $B>0$ for $\theta_{SH}=0.2$ in the absence of DMI ($D=0$). In this case, the reversal occurs via quasi-uniform magnetization precession, and therefore, the SDM reproduces accurately the magnetization dynamics (solid red line in Fig. \ref{fig:Fig1}(f)) computed from a full $\mu M$ point of view (black dots in Fig. \ref{fig:Fig1}(f)), confirming the validity of the uniform magnetization approach in the absence of DMI ($D=0$).  

The SDM stability phase diagrams showing the terminal out-of-plane magnetization direction as function of $B$ and $j(t)$ (with $t_{R}=t_{F}=200\mathrm{ps}$, $\tau=20\mathrm{ps}$ and different amplitudes $j_{a}$) are depicted in Fig. \ref{fig:Fig1}(g) and (h) for a high $\theta_{SH}=0.2$ and a more realistic $\theta_{SH}=0.11$ value of spin Hall angle respectively. These results were computed at room temperature by averaging over $10$ stochastic realizations. The same results were also obtained at zero temperature (see open circles in Fig. \ref{fig:Fig1}(g) and (h)). Note that $\theta_{SH}=0.2$ is around twice the value experimentally deduced for the Pt/Co from efficiency measurements\cite{Garello_14}, where $\theta_{SH}$ was estimated $0.11$. Therefore, these experiments\cite{Garello_14} cannot be reproduced by the SDM unless unrealistic values of $\theta_{SH}\sim 0.4$ are assumed\cite{Garello_14}. As it will be shown later, the key ingredient to achieve quantitative agreement is the presence of DMI, which can only be taken into account in a full $\mu M$ analysis.   

\textbf{Non-uniform magnetization patterns and current induced magnetization switching (CIMS) in the presence of finite DMI: micromagnetic results.} Although the SDM could qualitatively describe the stability phase diagrams, it fails to provide a quantitative description of the experiments\cite{Liu_12a,Garello_14}, and the spatial magnetization dependence ($\vec{m}(\vec{r},t)$) needs to be taken into account for a realistic analysis. Indeed, it has been argued that the Dzyaloshinskii-Moriya interaction (DMI) arises at the interface between the HM (Pt) and the FM (Co) layers\cite{Emori_13,Pizzini_14}. In particular, it was confirmed that apart from SL-SOT due to the SHE, also the DMI is a key ingredient in governing the statics and dynamics of DWs along ultrathin FM strips sandwiched in asymmetric stacks\cite{Thiaville_12,Emori_13,Ryu_13}. Similarly to the conventional symmetric exchange interaction ($\vec{H}_{exch}$) responsible of the ferromagnetic order, the interfacial DMI effective field $\vec{H}_{DMI}$ is only different from zero if the magnetization is a non-uniform continuous vectorial function $\vec{m}=\vec{m}(\vec{r},t)$. Apart from promoting non-uniform magnetization textures of a definite chirality in the bulk of the FM, the interfacial DMI also imposes specific boundary conditions (DMI-BCs) at the surfaces/edges of the sample\cite{Rohart_13}. Indeed, for finite DMI ($D\neq 0$), the DMI-BCs ensure that the local magnetization at the edges rotates in a plane containing the edge surface normal $\vec{n}$, and therefore, in a finite-ferromagnetic dot the uniform state is never a solution, so the SDM does no longer apply. Further details of the $\mu M$ are given in Methods.

\begin{figure}
\includegraphics[width=0.60\textwidth]{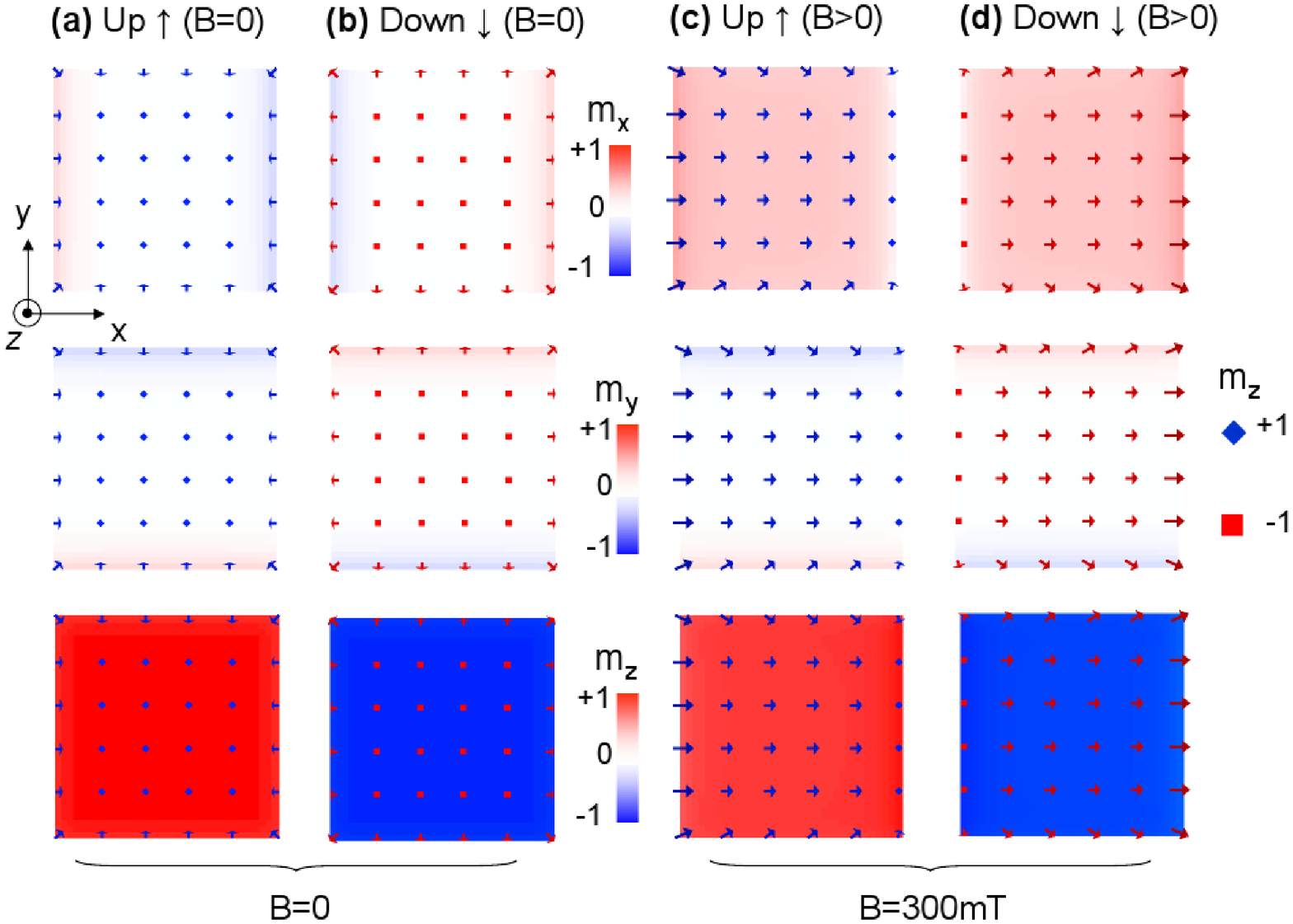}
\caption{\textbf{Non-uniform equilibrium magnetization patterns in the presence of finite DMI ($D=1.4\mathrm{mJ m^{-2}}$)} $|$ Magnetization snapshots depict the deviations of the local magnetization $\vec{m}(x,y)$ from the perfect out-of-plane direction as due to the DMI-BCs (equation (\ref{eq:DMI_BC})) at rest ($B=j=0$) in the presence of interfacial DMI ($D=1.4\mathrm{mJ/m^{2}}$) for an state mainly up magnetized ($<\vec{m}>\approx +\vec{u}_{z}$, $\uparrow$) (a), and for an state mainly down magnetized ($<\vec{m}>\approx -\vec{u}_{z}$, $\downarrow$) (b). Density plots of the longitudinal $m_{x}(x,y)$, transverse $m_{y}(x,y)$ and out-of-plane $m_{z}(x,y)$ configuration are shown from top to bottom respectively. Arrows show $m_{z}(x,y)$. (c) and (d) show the equilibrium state under a positive longitudinal field $B=+300\mathrm{nm}$ for the $\uparrow$ and $\downarrow$ states respectively.} 
\label{fig:Fig2}
\end{figure}

\textbf{Non-uniform equilibrium states under $B=0$ and $B\neq 0$ in the absence of current.} In the equilibrium state at rest ($B=j=0$), the average magnetization ($<\vec{m}(\vec{r})>$, where $<\ldots>$ represents the average in the FM volume) points mainly along the easy axis, either along $+\vec{u}_{z}$ ($\uparrow$, Fig. \ref{fig:Fig2}(a)) or $-\vec{u}_{z}$ ($\downarrow$, Fig. \ref{fig:Fig2}(b)). However, $\vec{m}(\vec{r})$ deviates from this easy axis direction at the edges (see Fig. \ref{fig:Fig2}). For the up $\uparrow$ state ($<\vec{m}>\approx +\vec{u}_{z}$), the local magnetization $\vec{m}(x,y)$ depicts a finite longitudinal component ($m_{x}\neq 0$), with $\vec{m}(0,y)=+|m_{x}|\vec{u}_{x}+m_{z}\vec{u}_{z}$ and $\vec{m}(L,y)=-|m_{x}|\vec{u}_{x}+m_{z}\vec{u}_{z}$ at the left $(0,y)$ and at the right $(L,y)$ laterals respectively (see Fig. \ref{fig:Fig2}(a)). Similarly, $\vec{m}(x,y)$ has a non-zero transversal component ($m_{y}\neq 0$), with $\vec{m}(x,0)=+|m_{y}|\vec{u}_{y}+m_{z}\vec{u}_{z}$ and with $\vec{m}(x,L)=-|m_{y}|\vec{u}_{y}+m_{z}\vec{u}_{z}$ at the bottom $(x,0)$ and top $(x,L)$ edges respectively. Instead of pointing inwards (Fig. \ref{fig:Fig2}(a)), the directions of the in-plane components $(m_{x},m_{y})$ at the edges reverse to outwards for the $\downarrow$ state ($<\vec{m}> \approx -\vec{u}_{z}$, Fig. \ref{fig:Fig2}(b)). The deviations from the perfect out-of-plane state are maximum at the edges and decrease over a distance given by $\sim \frac{2A}{D}$ toward to the sample center.

A moderate positive longitudinal field $\vec{B}=B\vec{u}_{x}$ well below the in-plane saturating field slightly modifies the out-of-plane magnetization in the central part of the FM sample, but it introduces significant changes in the local magnetization at the edges, as it can be seen in Fig. \ref{fig:Fig2}(c)-(d). A finite longitudinal component $m_{x}$ parallel to $\vec{B}$ arises at both bottom and top transverse edges $(y:0,L)$ (see $m_{y}(x,y)$ in Fig. \ref{fig:Fig2}(c)-(d)). Importantly, the effect of the positive field $\vec{B}=B\vec{u}_{x}$ with $B>0$ is opposite at the longitudinal left $(0,y)$ and right $(L,y)$ edges. Whereas $B>0$ supports the positive longitudinal magnetization component at the left edge $(0,y)$, it acts against the negative longitudinal magnetization component at the right edge $(L,y)$ for the $\uparrow$ state, as it is clearly seen in Fig. \ref{fig:Fig2}(c). For the $\downarrow$ state, $B>0$ supports the positive $m_{x}(L,y)$ and acts against the negative $m_{x}(0,y)$ (Fig. \ref{fig:Fig2}(d)).

\begin{figure}
\includegraphics[width=0.60\textwidth]{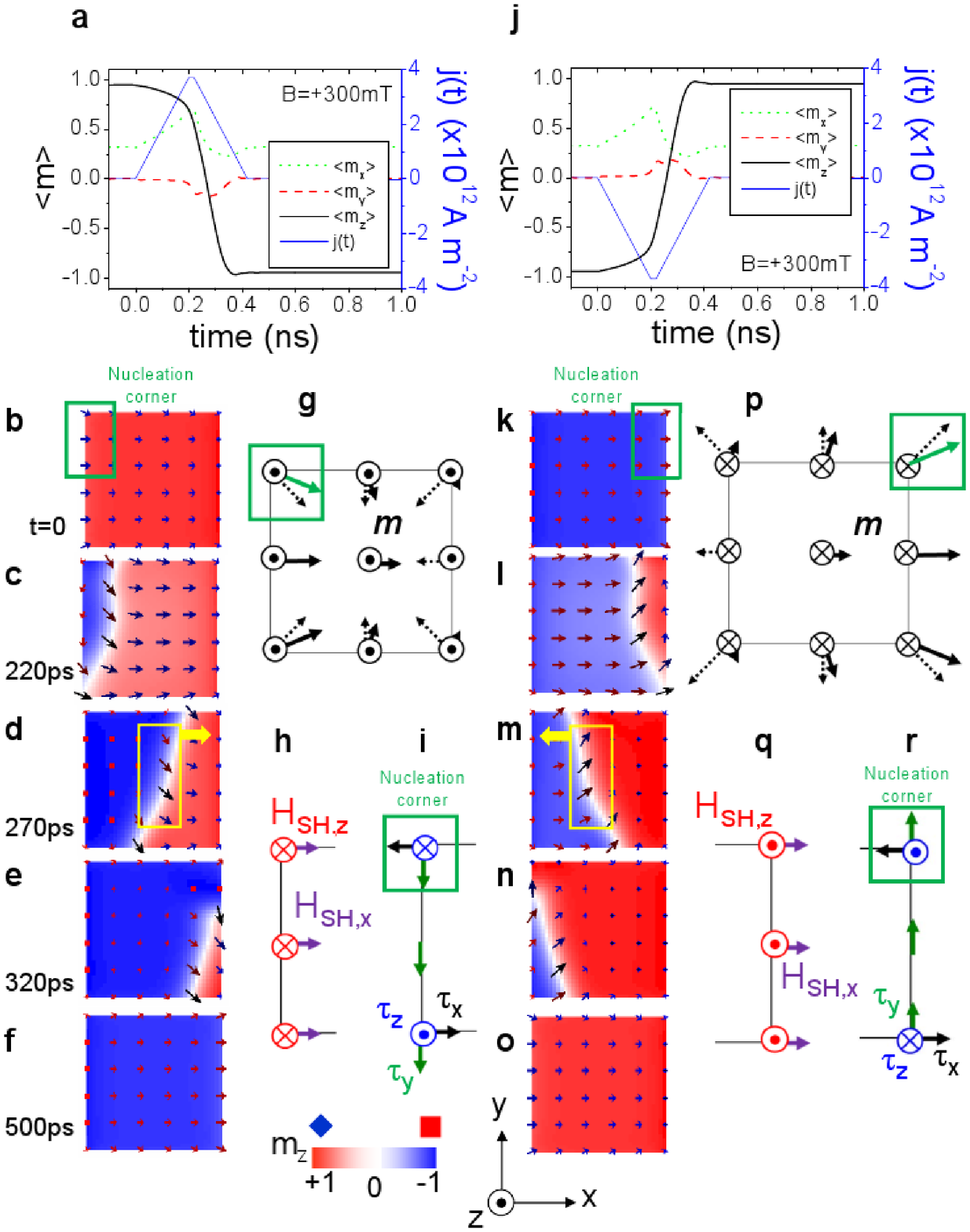}
\caption{\textbf{Non-uniform current-induced magnetization switching (CIMS) in the presence of DMI ($D=1.4\mathrm{mJ m^{-2}}$)} $|$ Graphs at the left panel correspond to the $\uparrow$ to $\downarrow$ switching under a positive current pulse ($j>0$). (a) Temporal evolution of the Cartesian components of the magnetization averaged over volume sample ($<m>$) for $B>0$ and $j_{a}>0$. The applied pulse $j(t)$ is also shown. (b)-(f) Magnetization $\vec{m}(x,y)$ snapshots during the $\uparrow$ to $\downarrow$ CIMS. Green box in (b) indicates the corner where the switching is triggered as explained in the text and in the schemes (g)-(i): (g) shows $\vec{m}(x,y)$ at different points of relevance to understand the CIMS. Dotted arrows indicate the in-plane components of the equilibrium $\vec{m}$ for $B=j=0$, whereas solid vectors indicate the equilibrium state under $B>0$. $B$ supports the in-plane longitudinal component at the left edge $(x,y)=(0,y)$. (h) Scheme of the out-of-plane component ($H_{SH,z}$, in red) and the in-plane longitudinal component ($H_{SH,x}$, in purple) of the SHE effective field ($\vec{H}_{SH}(x,y)$) at the left edge corresponding to (b) and (g). (i) Cartesian components of the local torque ($\tau_{x},\tau_{y},\tau_{z}$) due to $\vec{B}=B\vec{u}_{x}$ and $\vec{H}_{SH}(x,y)$ at the relevant left edge: the CIMS is triggered at the top left corner, where $\tau_{z}<0$ is opposed to the initial out-of-plane up magnetization. Graphs at the right panel (j)-(r) correspond to the $\downarrow$ to $\uparrow$ CIMS under $B>0$ but $j<0$. Yellow boxes in (d) and (m) indicate the internal structure of the current-driven domain wall motion due to the SHE.} 
\label{fig:Fig3}
\end{figure}

\textbf{Non-uniform CIMS from $\uparrow$ to $\downarrow$ with $B>0$ and $j(t)>0$ for finite DMI ($D\neq 0$).} Since for finite DMI ($D\neq 0$) the equilibrium states of Fig. \ref{fig:Fig2} depict non-uniform magnetization patterns $\vec{m}(\vec{x,y})$, and the SHE effective field depends on the local magnetization ($\vec{H}_{SH}(x,y)=H_{SH}\vec{m}(x,y)\times \vec{u}_{y}$), the magnetization dynamics must be also non-uniform, even for the small nano-sized confined dots with $L=90\mathrm{nm}$ with strong DMI. The non-uniform magnetization dynamics under static longitudinal field ($B=\pm 300\mathrm{mT}$) was studied under injection of current pulses  $\vec{j}=j(t)\vec{u}_{x}$ (Fig. \ref{fig:Fig1}(b)) with $t_{R}=t_{F}=200\mathrm{ps}$, $\tau=20\mathrm{ps}$ and $j_{a}=\pm 3.7\times 10^{12}\mathrm{A/m^{2}}$ (corresponding to an uniform current $I=1.2\mathrm{mA}$ through the Pt/Co section, $3.6\times 90 \mathrm{nm^{2}}$) by $\mu M$ solving the dynamics equation (Methods). The value for the spin Hall angle is $\theta_{SH}=0.11$ as deduced experimentally by Garello et al.\cite{Garello_14} for similar samples. The temporal evolution of the Cartesian magnetization components averaged over the volume of the FM ($<m_{i}>(t)$ with $i:x,y,z$) and the current pulse temporal profile ($j(t)$) are shown in Fig. \ref{fig:Fig3} for different combinations of $B$ and $j_{a}$ which promote the CIMS from $\uparrow$ to $\downarrow$ ($B>0$ and $j>0$), and from $\downarrow$ to $\uparrow$ ($B>0$ and $j<0$). Representative transient magnetization snapshots during the CIMS are also shown in Fig. \ref{fig:Fig3}, which clearly indicate that the switching is non-uniform as opposed to SDM predictions.

We focus our attention on the CIMS from $\uparrow$ to $\downarrow$ with $B>0$ and $j(t)>0$ (left graphs in Fig. \ref{fig:Fig3}) in the presence of strong DMI ($D=1.4\mathrm{mJ/m^{2}}$). The temporal evolution of the Cartesian magnetization components over the ferromagnet volume ($<m_{i}>$ with $i:x,y,z$) is shown in Fig. \ref{fig:Fig3}(a), whereas representative transient magnetization snapshots are shown in Fig. \ref{fig:Fig3}(b)-(f). The reversal takes place in two stages. The first one consists on the magnetization reversal at the top left corner of the square resulting in DW nucleation, and the second one occurs via current-driven domain wall (DW) propagation from the left to right due to the SHE. Apart from the snapshots of Fig. \ref{fig:Fig3}(b)-(f), these two stages are also evident in the temporal evolution of the out-of-plane magnetization $<m_{z}>$ shown in Fig. \ref{fig:Fig3}(a). From $t=0$ to $t\approx 200\mathrm{ps}$, $<m_{z}>$ decreases gradually, whereas it decreases almost linearly from $t\approx 200\mathrm{ps}$ to $t\approx 300\mathrm{ps}$, consistent with the current-driven DW propagation where its internal structure is seen in Fig. \ref{fig:Fig3}(d). 

The magnetization reversal during the first stage is non-uniform due to the DMI imposed boundary conditions (DMI-BCs, see Methods), but to understand in depth the underlaying reasons, it is needed to take into account the chiral-induced non-uniform magnetization ($\vec{m}(x,y)=m_{x}(x,y)\vec{u}_{x}+m_{y}(x,y)\vec{u}_{y}+m_{z}(x,y)\vec{u}_{z}$) in the presence of the applied field ($B>0$) and current ($j>0$). As it can be seen in Fig. \ref{fig:Fig2}(c) or in Fig. \ref{fig:Fig3}(b), $B>0$ and DMI-BCs support the positive longitudinal magnetization component ($m_{x}(x,y)$) at the left-edge $(x,y)=(0,y)$, whereas the negative $m_{x}(x,y)$ is very small at the right edge $(x,y)=(L,y)$. An schematic view of the local equilibrium magnetization at relevant locations is shown in Fig. \ref{fig:Fig3}(g) for the $\uparrow$ state under $B>0$ and zero current. The effective SHE field is also non-uniform: $\vec{H}_{SH}(x,y)=H_{SH,x}(x,y)\vec{u}_{x}+H_{SH,z}(x,y)\vec{u}_{z}$ with $H_{SH,x}(x,y)=-H_{SH}(j)m_{z}(x,y)$ and $H_{SH,z}(x,y)=H_{SH}(j)m_{x}(x,y)$. As the out-of-plane component $H_{SH,z}(x,y)$ is negative (note that $H_{SH}(j)<0$ for $j(t)>0$) and proportional to the local $m_{x}(x,y)$, which is maximum and positive at the left edge ($(x,y)=(0,y)$), the reversal starts from the left edge (see Fig. \ref{fig:Fig3}(h)). However, in addition to this asymmetry along the longitudinal $x$-axis imposed by the DMI-BCs and supported by $B$ (left $vs$ right edges), other chiral asymmetry arises along the transverse $y$-axis in the left edge: the reversal is first triggered from the top left corner ($(x,y)=(0,L)$), whereas the local CIMS is delayed at the bottom-left corner ($(x,y)=(0,L)$), as it clearly seen in Fig. \ref{fig:Fig3}(c). The reason for this transverse asymmetry relies in the different direction of local torque at the initial state (Fig. \ref{fig:Fig3}(i)). The relevant torque is the one experienced by the local magnetization at the left edge $\vec{m}(0,y)$ due to $\vec{H}_{SH}(x,y)$, which is also supported by $\vec{B}=\mu_{0}H_{x}\vec{u}_{x}$: $\vec{\tau}(x,y)=-\gamma_{0}\vec{m}(x,y)\times (\vec{H}_{SH}(x,y)+\vec{H}_{x})=\tau_{x}\vec{u}_{x}+\tau_{y}\vec{u}_{y}+\tau_{z}\vec{u}_{z}$. As the local transverse magnetization $m_{y}(x,y)$ has different sign at the top ($m_{y}(0,L)<0$) and bottom ($m_{y}(0,0)>0$) corners of the left edge, both the longitudinal component ($\tau_{x}(x,y)=-\gamma_{0}m_{y}(x,y)m_{x}(x,y)H_{SH}$) and the out-of-plane component of this torque ($\tau_{z}(x,y)=-\gamma_{0}(m_{y}(x,y)m_{z}(x,y)H_{SH}-m_{y}(x,y)H_{x})$) point in opposite directions at the top and the bottom corners of the left edge (see Fig. \ref{fig:Fig3}(i)). The relevant component of $\vec{\tau}(x,y)$ to understand the local reversal is the out-of-plane one: as  $\tau_{z}(0,L)<0$ at the top left corner but $\tau_{z}(0,0)>0$ at the bottom left corner, the reversal is firstly triggered from the top corner, where $\tau_{z}(0,L)$ opposes to the initial out-of-plane component of the magnetization ($\uparrow$). Once the local reversal is achieved at the top left corner, the switching expands from left to right and from top to bottom: the local in-plane magnetization at the bottom left edge rotates clockwise due to $H_{SH,z}(0,0)<0$, and once $m_{y}(0,0)$ becomes negative, also $\tau_{z}(0,0)<0$ promotes the local reversal. 

When all points at the left edge have reversed their initial out-of-plane magnetization ($m_{z}(0,y)<0$) a left-handed ($D>0$) down-up DW emerges, separating the reversed (with $\downarrow$) from the non-reversed (with $\uparrow$) zones. Note that once the local magnetization has reversed its initial out-of-plane direction, it experiences little torque due to $\vec{H}_{SH}$ (see Supplementary Information), so it is stable for the rest of the switching process, which takes place by current-driven DW propagation during the second stage. 

The internal structure of the propagating DW is shown in Fig. \ref{fig:Fig3}(d). Even in the presence of the longitudinal field ($B>0$), its internal moment ($\vec{m}_{DW}$) and its normal ($\vec{n}_{DW}$) do not point along the positive $x$-axis, and the DW depicts tilting or a rotation of its normal due to the SHE current-driven propagation. The DW tilting has been experimentally observed in the absence of in-plane field under high currents\cite{Ryu_12}, and theoretically studied, both in the absence and in the presence of in-plane fields, in elongated strips along the $x$-axis\cite{Emori_13b,Boulle_13,Martinez_14,Martinez_14b}. If the only driving force on the down-up DW ($\downarrow \uparrow $) were a strong positive (negative) current $j>0$ ($j<0$) with $B=0$, both $\vec{m}_{DW}$ and $\vec{n}_{DW}$ would rotate clockwise (counter-clock wise)\cite{Martinez_14}. Here, we observe that the DW tilting is also assisted during the DW nucleation due to the DMI-BCs, $\vec{B}$ and $\vec{H}_{SH}$. $B>0$ would support the internal longitudinal magnetization of the left-handed down-up DW if its normal points along the $x$-axis ($\vec{n}_{DW}=+\vec{u}_{x}$), as it would be the case of current-driven DW motion along an elongated strip along the $x$-axis\cite{Emori_13b}. However, due to the non-uniform local CIMS at the left edge in our confined dots, the DW normal has a non-zero negative transverse component ($n_{DW,y}<0$) for $B>0$ and $j>0$. As it is shown in the $270\mathrm{ps}$-snapshot of Fig. \ref{fig:Fig3}(d), in addition to a positive longitudinal component ($m_{DW,x}>0$), the internal DW moment also has a no-null negative transverse component ($m_{DW,y}<0$). Note that the direction of both $\vec{n}_{DW}$ and $\vec{m}_{DW}$ during the DW propagation is also the direction of the local magnetization at the top-left corner, where the reversal was initially launched (see Fig. \ref{fig:Fig3}(c),(i)).

The full magnetization switching is completed before the current pulse has been switched off (see Fig. \ref{fig:Fig3}(a)), when the propagating down-up DW ($\downarrow \uparrow$) reaches the right edge. Due to the DW tilting, the reversal occurs first at the top right corner ($(x,y)=(L,L)$) with respect to the bottom right corner ($(x,y)=(L,0)$) (see $320\mathrm{ps}$-snapshot of Fig. \ref{fig:Fig3}(e)). Although this second stage, consisting on current-driven DW propagation, is similar to the one already explained for elongated thin strips as driven by the SHE\cite{Martinez_14,Martinez_14b,Emori_13b}, the DW nucleation during the first stage has not been addressed so far for such small nano-sized confined dots, and as it was explained above it is mainly due to the longitudinal field $\vec{B}$ which supports the longitudinal magnetization component at the left edge imposed by the DMI-BCs.

\textbf{DISCUSSION}

\textbf{Universal chiral promoted current-induced magnetization switching (CIMS) in strong DMI systems.} The CIMS from $\uparrow$ to $\downarrow$ can also be achieved if both $B$ and $j$ reverse their directions ($B<0$ and $j<0$). As it is straightforwardly understood from the former description, in this case the reversal is triggered from the bottom right corner ($(x,y)=(L,0)$, where $\tau_{z}(L,0)<0$ opposes to the initial $\uparrow$ out-of-plane magnetization), and an up-down DW is driven toward the left (not shown). The CIMS from $\downarrow$ to $\uparrow$ under anti parallel field $B>0$ and current $j<0$ is shown at the right panel of Fig. \ref{fig:Fig3}(j)-(r).  

In general, the CIMS can be described as follows: ($\textbf{i}$) the initial out-of-plane magnetization direction ($\uparrow$ or $\downarrow$) determines the direction (inwards or outwards) of the local in-plane $\vec{m}$ at the edges imposed by the DMI-BCs. ($\textbf{ii}$) The longitudinal field $\vec{B}$ supports the longitudinal in-plane magnetization component ($m_{x}$) at one of the two lateral edges, and acts against it at the opposite one. ($\textbf{iii}$) For the favored lateral edge, the local magnetization reversal is triggered at the corner where the out-of-plane torque $\tau_{z}$ due to $\vec{H}_{SH}$ and $\vec{B}$ opposes to the initial out-of-plane magnetization component ($m_{z}$). After that, the reversal also takes place in the middle part of the selected edge, and finally, the other corner is also dragged into the reversed region with the formation of a tilted DW. ($\textbf{iv}$) The CIMS is completed by the current-driven DW propagation. 

Also remarkable is the fact that for the same current pulses as in Fig. \ref{fig:Fig3} the CIMS is not achieved in the framework of the SDM if a realistic value for the spin Hall angle is adopted ($\theta_{SH}=+0.11$)\cite{Garello_14}, and the same limitation was also observed by full $\mu M$ simulations in the absence of the DMI ($D=0$). All these simulations point out that, even for the small confined dots considered here ($L=90\mathrm{nm}$), the strong DMI and the BCs imposed by it are essential to describe the CIMS driven by the SHE from both quantitative and qualitative points of view. The DMI-triggered switching ($D\neq 0$) was also studied for other ultrathin ($L_{z}=0.6\mathrm{nm}$) squares ($L_{x}=L_{y}=L$) with different in-plane dimensions ($10\mathrm{nm}\le L \le 300\mathrm{nm}$) and reversal mechanism remains similar to the one already described and depicted in Fig.\ref{fig:Fig3}. Note that the smallest evaluated side ($L=10\mathrm{nm}$) is small than the minimum side required to achieve thermal stability ($L_{min} \approx 25\mathrm{nm}$) according to the conventional criterion: in order to maintain sufficient stability of the data storage over at least five years, the effective energy barrier given by $E_{b}=KV$ (with $K\approx 4.35 \times 10^{5}\mathrm{J/m^{3}}$ the effective uniaxial anisotropy constant from Ref. \cite{Garello_14}, and $V=L^{2}L_{z}$  the volume of the sample) should be larger than $ 55 K_{B}T$, where and $K_{B}$ Boltzmann constant. The reversal was also similar under realistic conditions including disorder due to the edge roughness and thermal effects (see Supplementary Information). Moreover, this chiral CIMS, either from $\uparrow$ to $\downarrow$ or from $\downarrow$ to $\uparrow$, does not change when the FM Co layer is patterned with a disk shape (see Supplementary Information). It was also verified that this non-uniform reversal mechanism, consisting on DW nucleation and propagation, does not depend on the specific temporal profile of the applied pulse, provided its magnitude ($j_{a}$) and duration ($\tau$) are sufficient to promote the complete reversal for each $L$.

\begin{figure}
\includegraphics[width=0.70\textwidth]{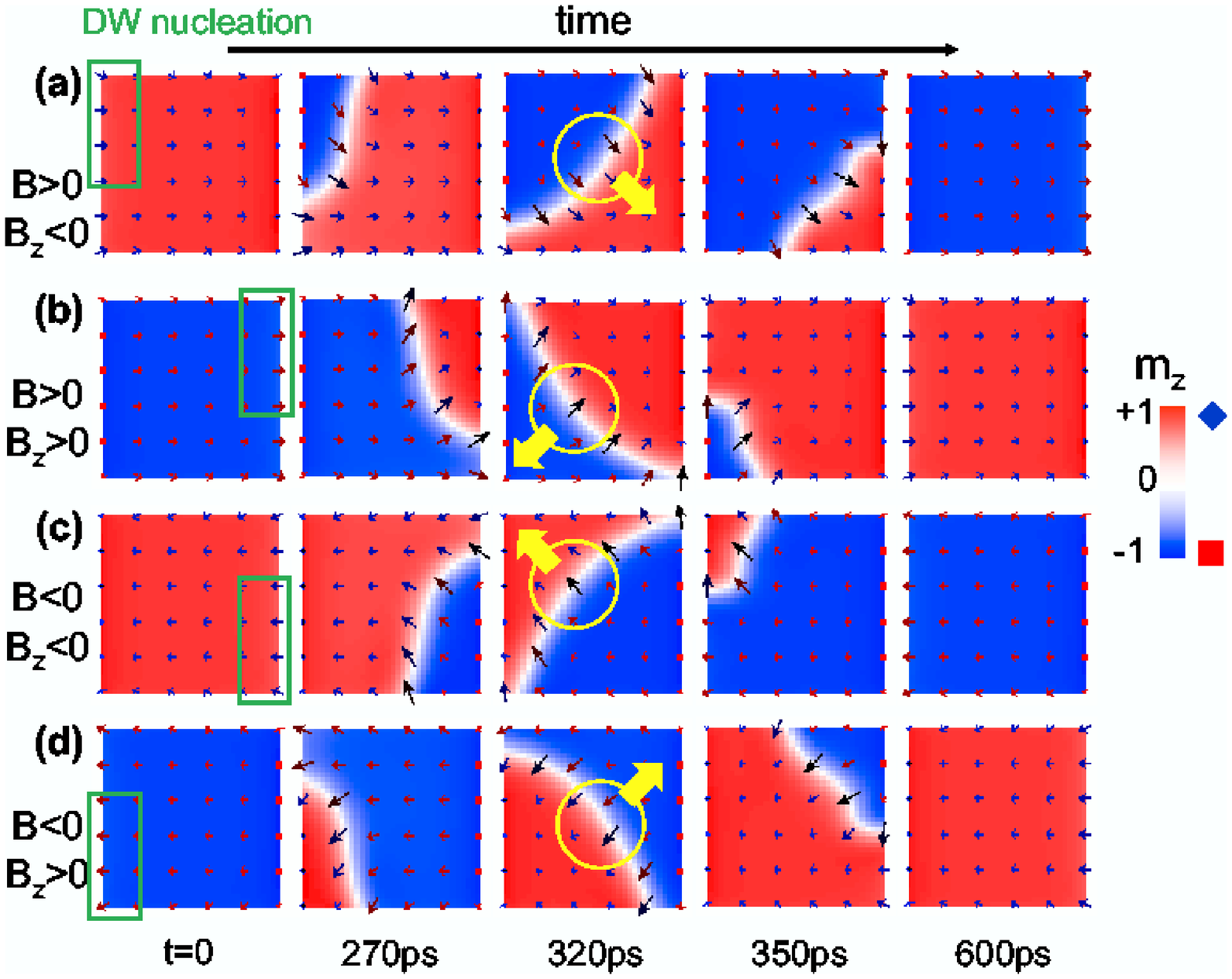}
\caption{\textbf{Field-induced magnetization switching in a ultrathin ($L_{z}=0.6\mathrm{nm}$) square dot with $L=90\mathrm{nm}$} $|$ Transient snapshots magnetization $\vec{m}(\vec{r})$ in the presence of interfacial DMI ($D=1.4\mathrm{mJ/m^{2}}$) during the reversal for different combinations of $B$ and $B_{z}$ with $|B|=300\mathrm{mT}$, $|B_{z}|=310\mathrm{mT}$, $t_{R}=t_{F}=200\mathrm{ps}$ and $\tau=20\mathrm{ps}$. The up ($\uparrow$) to down ($\downarrow$) switching is shown for ($B>0$,$B_{z}<0$) and ($B<0$,$B_{z}<0$) in (a) and (c) panels respectively, whereas the down ($\downarrow$) to up ($\uparrow$) is shown in (b) and (d) for ($B>0$,$B_{z}>0$) and ($B<0$,$B_{z}>0$). Green boxes indicate the region where the DW nucleation starts for each combination of initial state ($\uparrow$ or $\downarrow$), $B$ and $j$. Yellow circles indicate the field-driven propagating DW (yellow arrow indicate the direction of the reversed domain expansion). Note that the in-plane components of the nucleation region (green box) point along close to the internal DW moment (yellow circle) during its propagation.} 
\label{fig:Fig4}
\end{figure}

\textbf{Chiral nature of the field-induced magnetization switching (FIMS).} An analogous CIMS mechanism to the one described here for nano-size samples ($\approx 90\mathrm{nm}\times 90\mathrm{nm}$) was recently observed by Yu et al.\cite{Yu_14} using Kerr microscopy for an extended Ta($5\mathrm{nm}$)/CoFeB($1\mathrm{nm}$)/TaO($1.5\mathrm{nm}$) stack with micro-size in-plane dimensions ($20\mathrm{\mu m}\times 130\mathrm{\mu m}$). In that work, right-handed DWs ($D<0$) were nucleated assisted by the in-plane field and displaced along the current direction due to the negative spin Hall angle of the Ta. More recently, Pizzini et al.\cite{Pizzini_14} also used Kerr microscopy to visualize the asymmetric chiral DW nucleation under in-plane field and its subsequent propagation along extended ($\approx 70\mathrm{\mu m}$) Pt($3\mathrm{nm}$)/Co($0.6\mathrm{nm}$)/AlO($2\mathrm{nm}$) thin-films driven by out-of-plane field ($\vec{B}_{oop}=B_{z}\vec{u}_{z}$). Similar to our study, starting from the up state ($\uparrow$), a positive (negative) in-plane field ($\vec{B}=B\vec{u}_{x}$) promotes the local magnetization reversal at the left (right) edge, which was propagated to the right (left) driven by a negative out-of-plane field $B_{z}<0$. Their images indicate the nucleated DW has a left-handed chirality and it propagates without significant tilting due to the extended unconfined in-plane dimensions ($L_{y}\approx 70\mathrm{\mu m}$). In order to understand these observations, the field-induced magnetization switching (FIMS) has been also studied for confined small squares with $L=90\mathrm{nm}$ (the same geometry as in the former CIMS analysis) and others with lateral dimensions one order of magnitude larger ($L=1000\mathrm{nm}$). Static longitudinal fields $\vec{B}=B\vec{u}_{x}$ with $B>0$ and $B<0$ ($|B|=300\mathrm{mT}$) are applied along with short out-of-plane field pulses with $\vec{B}_{oop}(t)=B_{z}(t)\vec{u}_{z}$ with $|B_{z}|=310\mathrm{mT}$, $t_{R}=t_{F}=200\mathrm{ps}$ and $\tau=20\mathrm{ps}$ (the temporal profile of this pulse is the same as for the current-induced magnetization switching). The results for the confined $L=90\mathrm{nm}$ square dot are shown in Fig. \ref{fig:Fig4} for different combinations of the initial state ($\uparrow$ and $\downarrow$), in-plane static field $B$ ($B>0$ and $B<0$) and out-of-plane field pulse ($B_{z}(t)>0$ and $B_{z}(t)<0$). Similarly to the CIMS, the FIMS starts from an edge selected by the direction of $B$, with an even more evident chiral asymmetry between the two corners. Note again that the corner where the reversal starts has a transverse magnetization component ($m_{y}(x,y)$) pointing in the same direction as the transverse internal magnetization of the nucleated DW. Once the local switching has been triggered, the reverse domain (pointing along the opposite $z$ direction with respect to the initial state) expands asymmetrically along the longitudinal ($x$) and transverse ($y$) directions (see for instance snapshots at $t=270\mathrm{ps}$ and $t=320\mathrm{ps}$ in Fig. \ref{fig:Fig4}). Although here just a quarter-of-bubble is developed due to the confined shape at the corner, this asymmetric field-driven chiral expansion is similar to the one recently observed\cite{Je_13,Hrabec_14} in extended thin films. Moreover, our study also points out a qualitative difference between the current-driven and the field-driven nucleation: while the first one is driven by a non-uniform SHE out-of-plane effective field ($B_{SH,z}(x,y)=\mu_{0}H_{SH,z}(x,y)$ which depends on local $m_{x}(x,y)$), the second one is promoted by a uniform out-of-plane field $B_{z}$. Therefore, the current-induced nucleated DW propagates along the current direction ($x$-axis, see yellow arrows in Fig. \ref{fig:Fig3}(d) and (m)), whereas the field-driven DW expands radially from the corner (see yellow arrows in Fig. \ref{fig:Fig4}). Nevertheless, the fact that similar chiral local magnetization reversal occurs also at the corners of nano-size confined dots ($\approx 100\mathrm{nm}$ and below) clearly confirms the universality of the chiral reversal mechanism in these nano-size confined dots with strong DMI.

On the other hand, the Kerr images by Pizzini et al.\cite{Pizzini_14} do not show the corners of their extended thin-film (which is unconfined along the transverse $y$-axis) which are precisely where our modeling points out additional chiral asymmetry in the DW nucleation for confined dots (Fig. \ref{fig:Fig4}). Moreover, in their thin-films the field-driven DW does not depict tilting. In order to contrast these observations with our $\mu M$ predictions, the field-driven nucleation and propagation in an confined square dot has been also analyzed here, but with lateral in-plane dimensions one order of magnitude larger ($L=1000\mathrm{nm}$). We note that as $L$ is increased to the microscale, the nucleated DW is almost straight, with its normal oriented along the $x$-axis (no DW tilting), in the middle part of the nucleating edge (far form the corners). However, an asymmetry between the top and bottom corners is still present even for $L=1000\mathrm{nm}$ (see Supplementary Information): the reversal from $\uparrow$ to $\downarrow$ (from $\downarrow$ to $\uparrow$) is anticipated at the top-left (top-right) corner with respect to the bottom one under $B>0$ and $B_{z}<0$ ($B<0$ and $B_{z}<0$). This chiral asymmetry at the corners of the extended micro-size sample is similar to the observed for a confined dot (see. Fig. \ref{fig:Fig4}), and although it has not been addressed before, it could be observed by high resolution techniques\cite{Tetienne_14}.

\begin{figure}
\includegraphics[width=0.790\textwidth]{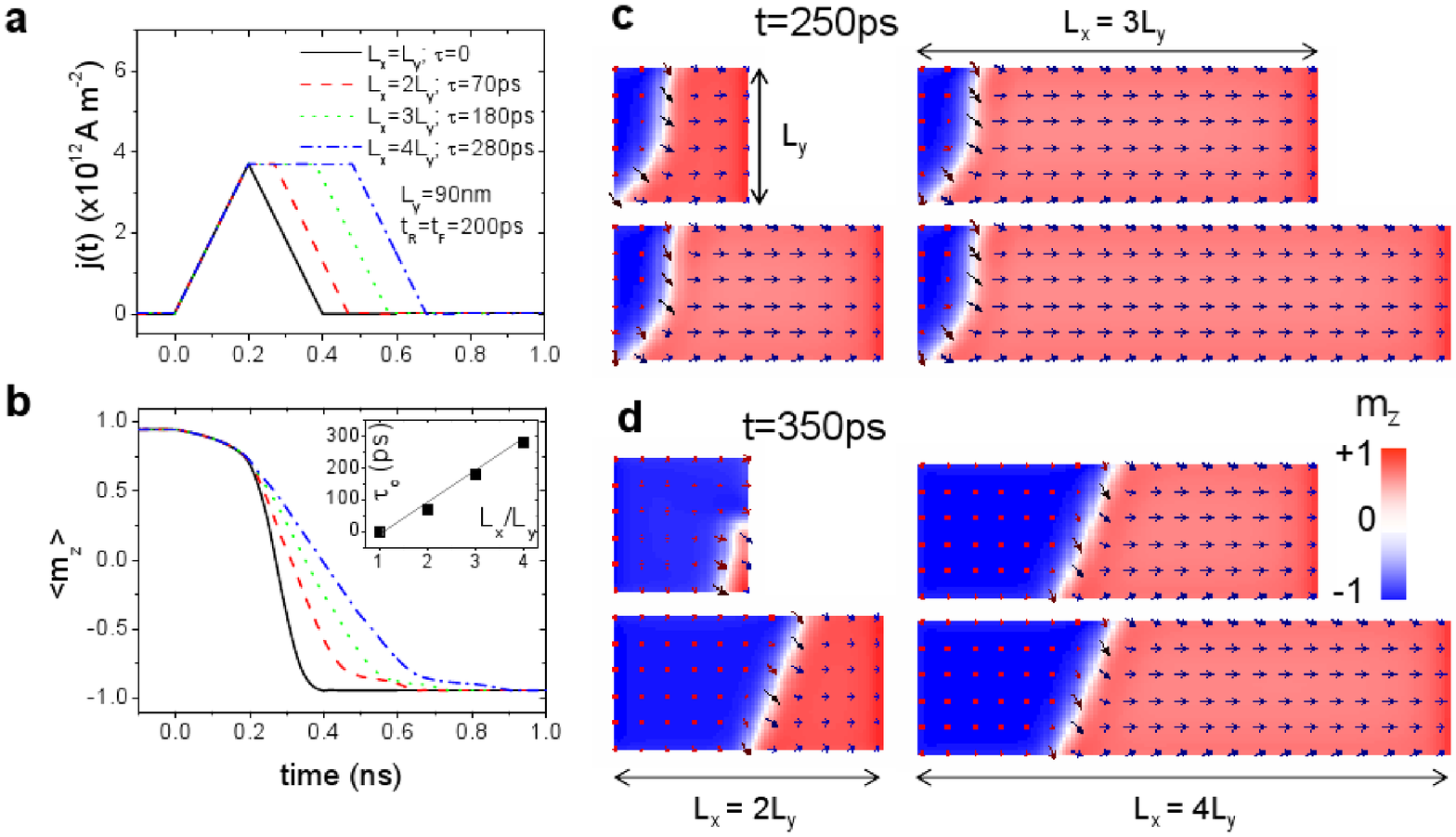}
\caption{\textbf{Current-induced magnetization switching along thin rectangles} $|$ CIMS in rectangles with $L_{y}=90\mathrm{nm}$ and different aspect-ratio $L_{x}/L_{y}:1,2,3,4$. The applied field $B=+300\mathrm{mT}$ and the current pulses have $j_{a}=3.7\times 10^{12}\mathrm{A/m^{2}}$ and $t_{R}=t_{F}=200\mathrm{ps}$ fixed, and different durations $\tau$ depending on $L_{x}/L_{y}$ (a). (b) Temporal evolution of the out-of-plane component $<m_{z}>$ for different rectangles under the pulses shown in (b). Snapshots of the magnetization state at $t\approx 250\mathrm{ps}$ (c) (DW nucleation) and at $t\approx 350\mathrm{ps}$ (d). The inset in (b) shows the critical threshold for $\tau$ as a function of the $L_{x}/L_{y}$.}
\label{fig:Fig5}
\end{figure}

\textbf{CIMS in confined nanodots with rectangular shape.} The CIMS was also studied in rectangles with different in-plane aspect-ratios $1\leq L_{x}/L_{y}\leq 4$ (Fig. \ref{fig:Fig5}(a)-(c)). The thickness is fixed ($L_{z}=0.6\mathrm{nm}$) as before. Again the switching takes place by DW nucleation followed by its current-driven propagation along the $x$-axis, which further supports the universality of the reversal mechanism in systems with strong DMI. In this case, the nucleation takes place during the first $200\mathrm{ps}$ independently of the rectangle aspect-ratio $L_{x}/L_{y}$, but the critical pulse duration ($\tau$) for fixed $j_{a}$ and $t_{R}=t_{F}$, increases linearly with $L_{x}/L_{y}$ (see the inset in Fig. \ref{fig:Fig5}(c)), an prediction which could be experimentally validated to estimate both the spin Hall angle ($\theta_{SH}$) and the DMI parameter ($D$) if the rest of material parameters ($M_{s}$, $A$, $K_{u}$, $\alpha$) are known by other means.

\begin{figure}
\includegraphics[width=0.50\textwidth]{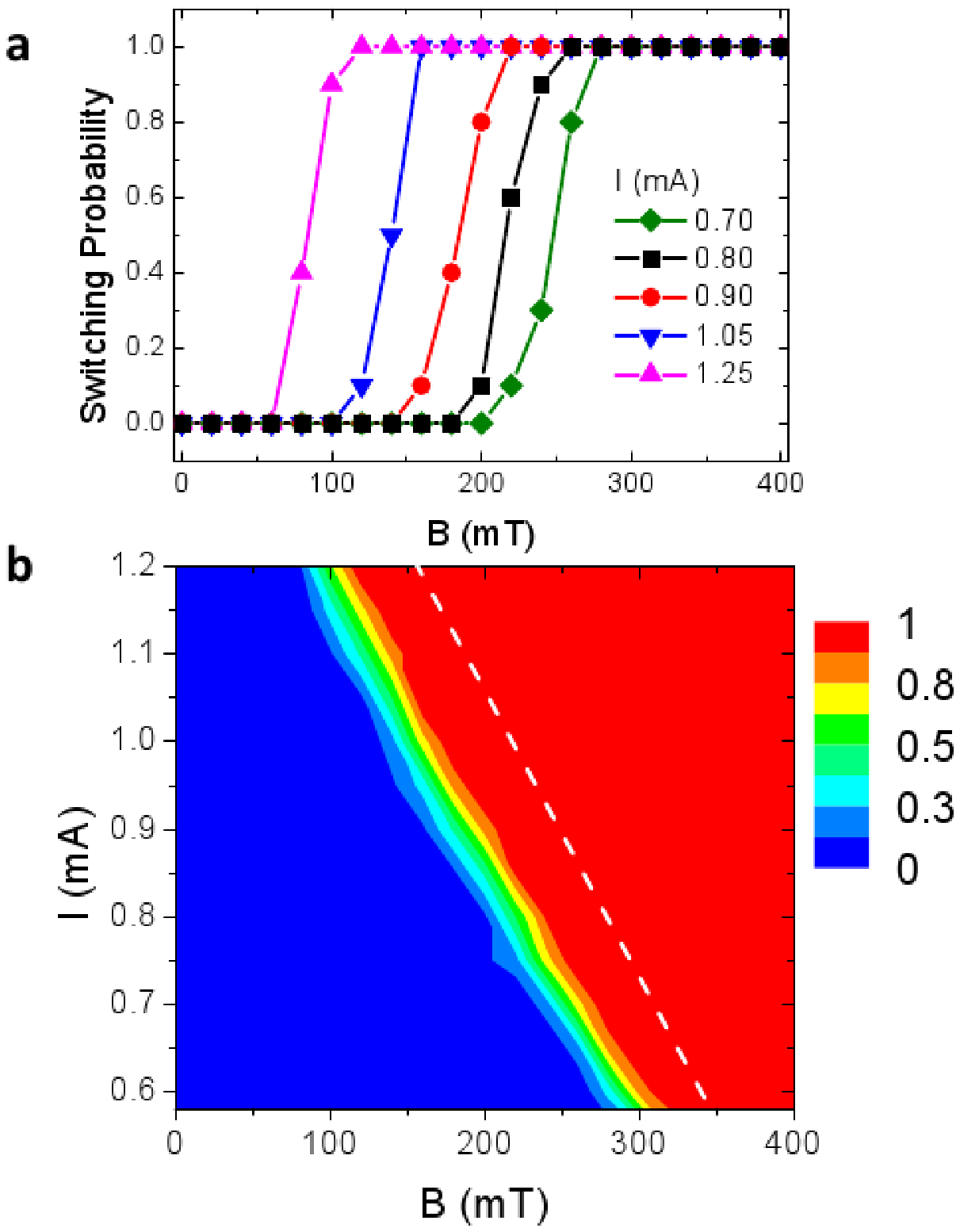}
\caption{\textbf{Quantitative description of experimental results} $|$ Micromagnetically computed switching probability as function of the applied field $B$ and the current pulse $j(t)$. Similar current pulses as in the experiments by Garello et al. \cite{Garello_14} are applied: $t_{R}=t_{F}=200\mathrm{ps}$ and $\tau=20\mathrm{ps}$ are fixed, and different magnitudes $j_{a}=I/(3.6\times 90\mathrm{nm^{2}})$ are studied ($I=1.2\mathrm{mA}$ corresponds to $j_{a}=3.7\times 10^{12}\mathrm{A/m^{2}}$). Results were computed at room temperature $T=300\mathrm{K}$ by averaging over $10$ stochastic realizations. (a) Switching probability as a function of $B$ for pulses with several magnitudes expressed in term of $I$ as in the experimental study. The switching probability is as function $B$ and $I$ is depicted by density plot in (b). Dashed white curve in (b) represents the threshold between not-switching and switching computed at zero temperature.} 
\label{fig:Fig6}
\end{figure}

\textbf{Comparison to experiments of current-induced magnetization switching.} Although our study goes further than a mere comparison to available experimental results, it is interesting to show how the non-uniform CIMS can explain quantitatively the experimental measurements by considering realistic material parameters (see Methods and Supplementary Information). With the aim of providing an explanation of experimental observations\cite{Garello_14} for the ultrahin Co square with $L=90\mathrm{nm}$ in a Pt($3\mathrm{nm}$)/Co($0.6\mathrm{nm}$)/AlO($2\mathrm{nm}$) stack, we have repeated the former study for several values of the applied field ($B$) and different different magnitudes of the current pulse ($j_{a}$). The rise and fall times ($t_{R}=t_{F}=200\mathrm{ps}$) and the duration ($\tau=20\mathrm{ps}$) of the pulse were maintained fixed as in the experimental study \cite{Garello_14}. Here we consider the up state ($<m_{z}>\approx +1$, $\uparrow$) as the initial one. For each $(B,j_{a})$, the switching probability at room temperature was computed as the averaged over $10$ stochastic realizations. Realistic conditions were taken into account by considering random edge roughness with characteristic sizes ranging from $0.5\mathrm{nm}$-$5\mathrm{nm}$ (see Methods). The $\mu M$ results are collected in Fig. \ref{fig:Fig6} which indicates a good quantitative agreement with recent experimental measurements\cite{Garello_14}.

It was verified that the CIMS mechanism (local magnetization reversal with DW nucleation and subsequent current-driven propagation) remains qualitatively unchanged even under these realistic conditions (see Supplementary Information). Moreover, although marginal discrepancies between these $\mu M$ data (Fig. \ref{fig:Fig6}) and the experimental results shown in Fig. 2(d) of ref.\cite{Garello_14} can be seen, the quantitative agreement is remarkable considering similar material parameters as inferred experimentally\cite{Garello_14}: $M_{s}=8.7\times 10^{5}\mathrm{A/m}$, $A=1.6\times 10^{-11}\mathrm{J/m}$, $K_{u}=8\times 10^{5}\mathrm{J/m^{3}}$, $\alpha=0.3$, $\theta_{SH}=0.11$ and $D=1.4\mathrm{mJ/m^{2}}$ (see Supplementary Information for detailed justification of these inputs). Note that with the SDM a quantitative agreement with the experimental data was only achieved with unrealistic values of the ($\theta_{SH}=0.4$)\cite{Garello_14}. Note that the DMI parameter $D=1.4\mathrm{mJ/m^{2}}$ was not determined experimentally\cite{Garello_14}, but the fact that this value $D=1.4\mathrm{mJ/m^{2}}$ provides reasonable quantitative agreement with their experiments, and that this value is also in good quantitative agreement with very recent estimations by other means for similar Pt/Co/AlO systems\cite{Cho_14} constitute additional evidences that our modeling is compatible with the dominant physics behind these CIMS processes.

\textbf{Conclusions.} In summary, the current-driven magnetization switching in ultrathin HM/FM/Oxide heterostructures with high PMA and strong DMI has been studied by means of full micromagnetic simulations. Even for the small in-plane dimensions ($\sim 100\mathrm{nm}$), the analysis points out that the magnetization reversal mechanism is non-uniform. It starts by local magnetization reversal induced by the SHE and assisted by the in-plane field in collaboration with the DMI boundary conditions. The longitudinal field and the DMI imposed boundary conditions select the lateral/edge and the specific corner at which the nucleation is triggered, where the relevant torques due to the SHE and the longitudinal field accelerate the local reversal. After that, the switching is completed by current-driven domain wall propagation driven by the SHE, where the current direction determines the direction of the wall motion, and the internal magnetization of the propagating wall points closely to the local magnetization at the selected corner where the reversal was initially launched. Similar nucleation and propagation mechanisms were also observed under out-of-plane fields, confirming again the chiral-triggered magnetization reversal in these nano-size confined dots. These results clearly exclude the single domain approach as a proper model to describe these switching experiments, and therefore, the estimations of the spin Hall angle based in this oversimplified model should be revised by adopting a much more realistic full 3D micromagnetic approach. Moreover, by analyzing the switching under realistic conditions including disorder and thermal effects, it was found that the mechanism is universal, and for instance, it could be used to the quantify both the DMI and the spin Hall angle by studying the reversal of ferromagnetic layers with different length for fixed width and thickness. As the reversal mechanism occurs in a reliable and efficient way, and more importantly, as it is also highly insensitive to defects and thermal fluctuations, our results are also very relevant for technological recording applications combining non-volatility, high stability, ultra-dense storage and ultrafast writing.

\textbf{METHODS}

\textbf{Magnetization dynamics under SOT due to the SHE.} Under injection of a spatially uniform current density pulse along the $x$-axis $\vec{j}(t)=j(t)\vec{u}_{x}$ (see its temporal profile in Fig. \ref{fig:Fig1}(b)), the magnetization dynamics is governed by the augmented Landau-Lifshitz Gilbert eq.

\begin{equation}
\frac{d\vec{m}}{dt} = -\gamma_{0}\vec{m}\times \left( \vec{H}_{eff} + \vec{H}_{th} \right)+
         \alpha\left(\vec{m}\times \frac{d\vec{m}}{dt}\right) -\gamma_{0}H_{SH}\vec{m}\times(\vec{m}\times\vec{\sigma}) \label{eq:LLG_SHE}
\end{equation}

\noindent where $\vec{m}(\vec{r},t)=\vec{M}(\vec{r},t)/M_{s}$ is the normalized local magnetization with $M_{s}$ saturation magnetization, $\gamma_{0}$ is the gyromagnetic ratio and $\vec{H}_{eff}$ is effective field derived from the energy density of system ($\vec{H}_{eff}=-\frac{1}{\mu_{0}M_{s}}\frac{\delta \epsilon}{\delta \vec{m}}$). The first term in equation (\ref{eq:LLG_SHE}) represents the precessional torque of $\vec{m}$ around $\vec{H}_{eff} + \vec{H}_{th}$, where $\vec{H}_{th}$ is the thermal field representing the effect of thermal fluctuations at finite temperature. $\vec{H}_{th}$ is a white-noise Gaussian-distributed stochastic random process with zero mean value (its statistical properties are given below). The second term in equation (\ref{eq:LLG_SHE}) is the damping torque with $\alpha$ the dimensionless Gilbert damping parameter. The last term in equation (\ref{eq:LLG_SHE}) is the SL-SOT from the spin Hall effect (SHE), where $\vec{\sigma}=\vec{u}_{y}$ is the unit vector pointing along the direction of spin current polarization due to the SHE in the Pt layer, and $H_{SH}$ represents the magnitude of the effective spin Hall field $\vec{H}_{SH}=H_{SH}\vec{m}\times\vec{\sigma}$ given by   

\begin{eqnarray}
H_{SH}(t) &=& \frac{\hbar\theta_{SH}j(t)}{2e\mu_{0}M_{s}L_{z}}   \label{eq:SpinHallField1}
\end{eqnarray}

\noindent where $L_{z}$ is thickness of the FM layer, $\hbar$ is Planck's constant, $e<0$ is the electron charge and $j(t)$ is the instantaneous value of the electrical density current. As in the experiment by Garello et al.\cite{Garello_14}, the current is assumed to flow uniformly through the HM/FM bilayer (see Supplementary Information for additional discussion). $\theta_{SH}$ is the Spin Hall angle, which is defined as the ratio between the spin and charge current densities.

\textbf{Single Domain Model (SDM).} If the magnetization is assumed to be spatially uniform ($\vec{m}(t)=m_{x}(t)\vec{u}_{x}+m_{y}(t)\vec{u}_{y}+m_{z}(t)\vec{u}_{z}$), the deterministic effective $H_{eff}$ field in equation (\ref{eq:LLG_SHE}) only includes the PMA anisotropy, magnetostatic and Zeeman contributions $\vec{H}_{eff}=\vec{H}_{PMA}+\vec{H}_{dmg}+\vec{H}_{ext}$. The Zeeman contribution due to the longitudinal field is $\vec{H}_{ext}=(B/\mu_{0})\vec{u}_{x}$. The uniaxial PMA anisotropy effective field is

\begin{eqnarray}
\vec{H}_{PMA} &=& \frac{2K_{u}}{\mu_{0}M_{s}}m_{z}\vec{u}_{z} \label{eq:PMAFieldSDM}
\end{eqnarray}

\noindent and the demagnetizing field in the SDM approach is expressed as

\begin{eqnarray}
\vec{H}_{dmg} &=& -M_{s} N \vec{m} = -M_{s} \left( N_{x}m_{x}\vec{u}_{x} + N_{y}m_{y}\vec{u}_{y} + N_{z}m_{z}\vec{u}_{z}\right) \label{eq:PMAFieldSDM}
\end{eqnarray}

\noindent where $N$ is the diagonal magnetostatic tensor with $N_{x}=N_{y}=0.011$ and $N_{z}=0.975$ being the self-magnetostatic factors\cite{Aharoni_98} for $L=90\mathrm{nm}$ and $L_{z}=0.6\mathrm{nm}$.

The thermal field $\vec{H}_{th}(t)$ is a stochastic vector process whose magnitude is related to the temperature $T$ via the fluctuation-dissipation theorem\cite{Brown_63}.   

\begin{eqnarray}
\vec{H}_{th}(t) &=& \vec{\eta}(t) \sqrt{\frac{2\alpha K_{B}T}{\gamma_{0}\mu_{0}M_{s}V\Delta t}} \label{eq:ThermalFieldSDM}
\end{eqnarray}

\noindent where $K_{B}$ is the Boltzmann constant, $V=L_{x}L_{y}L_{z}$ is the volume of the sample, $\Delta t$ is the time step, and $\vec{\eta}(t)=(\eta_{x},\eta_{y},\eta_{z})$ is a Gaussian distributed white-noise stochastic vector with zero mean value ($<\eta_{i}(t)>=0$ for $i:x,y,z$) and uncorrelated in time ($<\eta_{i}(t)\eta_{j}(t')>=\delta_{i,j}\delta(t-t')$, where $\delta_{i,j}$ is the Kronecker delta and $\delta(t-t')$ the Dirac delta). Here $<...>$ means the statistical average over different stochastic realizations of the stochastic process. Equation (\ref{eq:LLG_SHE}) was numerically solved with a $4^{th}$-order Runge-Kutta scheme with a time step of $0.1\mathrm{ps}$.  

\textbf{Micromagnetic Model ($\mu M$).} When the spatial dependence of the magnetization is taken into account ($\vec{m}(\vec{r},t)$), the deterministic effective field $\vec{H}_{eff}(\vec{r},t)$ in equation (\ref{eq:LLG_SHE}) includes the space-dependent exchange $\vec{H}_{exch}(\vec{r},t)=\frac{2A}{\mu_{0}M_{s}}(\nabla^{2}\vec{m})$ with $A$ the exchange constant, and the interfacial DMI $\vec{H}_{DMI}(\vec{r},t)=-\frac{2D}{\mu_{0}M_{s}}\left[ \left(\nabla \cdot \vec{m} \right) \vec{u}_{z} - \nabla m_{z} \right]$\cite{Thiaville_12,Rohart_13} where $D$ is a parameter describing the DMI magnitude. Both the local Zeeman and PMA uniaxial contributions to $\vec{H}_{eff}(\vec{r},t)$ are computed similarly as in the SDM ($\vec{H}_{ext}=\frac{B}{\mu_{0}}\vec{u}_{x}$ and $\vec{H}_{PMA}(\vec{r},t)=\frac{2K_{u}}{\mu_{0}M_{s}}m_{z}(\vec{r},t)\vec{u}_{z}$). Note also that in the $\mu M$ the magnetostatic field $\vec{H}_{dmg}(\vec{r},t)$ is also space-dependent on $\vec{m}(\vec{r},t)$ everywhere. The Oersted field due to the current was also taken into account but it was found irrelevant and very small as compared to the other dominant contributions in $\vec{H}_{eff}$. (see\cite{GPMagnet,LopezDiaz_12} for the numerical details).

In the absence of DMI ($D=0$), the symmetric exchange interaction imposes boundary conditions (BCs) at the surfaces of the sample\cite{Bertotti_98} so that $\vec{m}(\vec{r})$ does not change along the surface ($\partial \vec{m}/\partial n =0$, where $\partial/\partial n$ indicates the derivative in the outside direction normal to the surface of the sample). However, in the presence of the interfacial DMI ($D\neq 0$), these BCs have to be replaced by\cite{Rohart_13,Emori_13b}

\begin{equation}
\frac{\partial \vec{m}}{\partial n}= -\frac{D}{2A}\vec{m}\times (\vec{n}\times \vec{u}_{z})
\label{eq:DMI_BC} 
\end{equation}

\noindent where $\vec{n}$ represents the local unit vector normal to each sample surface. 

In the $\mu M$ the thermal field $\vec{H}_{th}(\vec{r},t)$ is also a stochastic vector process given by   

\begin{eqnarray}
\vec{H}_{th}(\vec{r},t) &=& \vec{\eta}(\vec{r},t) \sqrt{\frac{2\alpha K_{B}T}{\gamma_{0}\mu_{0}M_{s}\Delta V\Delta t}} \label{eq:ThermalFieldSDM}
\end{eqnarray}

\noindent where now $\Delta V=\Delta x\Delta y \Delta z$ is the volume of each computational cell and $\vec{\eta}(\vec{r},t)=(\eta_{x},\eta_{y},\eta_{z})$ is a white-noise Gaussian distributed stochastic vector with zero mean value ($<\eta_{i}(\vec{r},t)>=0$ for $i:x,y,z$) and uncorrelated both in time and in space ($<\eta_{i}(\vec{r},t)\eta_{j}(\vec{r'},t')>=\delta_{i,j}\delta(t-t')\delta(\vec{r}-\vec{r'})$). Most of the simulations for perfect samples were performed with a 2D discretization using cells of $\Delta x= \Delta y=2.5\mathrm{nm}$ in side, and thickness equal to the ferromagnetic layer ($L_{z}=0.6\mathrm{nm}$). Several tests were performed with cell sizes of $0.5\mathrm{nm}$ to confirm the numerical validity of the presented results. Realistic samples were also studied by considering edge roughness using cell sizes of $0.5\mathrm{nm}$. These realistic conditions are introduced by randomly generating edge roughness patterns with different characteristic sizes $0.5\mathrm{nm} \le D_{g} \le 5\mathrm{nm}$ at all edges. Equation (\ref{eq:LLG_SHE}) was numerically solved with a $6^{th}$-order Runge-Kutta scheme with a time step of $0.01\mathrm{ps}$ by using GPMagnet\cite{GPMagnet}, a commercial parallelized finite-difference micromagnetic solver\cite{LopezDiaz_12}.    

\textbf{Material parameters.} Typical high PMA material parameters were adopted for the results collected in the main text in agreement with experimental values for Pt/Co/AlO\cite{Pizzini_14,Garello_13,Garello_14}: saturation magnetization $M_{s}=8.7\times10^5\mathrm{A/m}$, exchange constant $A=1.6\times 10^{-11}\mathrm{J/m}$, uniaxial anisotropy constant $K_{u}=8.7\times10^{5}\mathrm{J/m^{3}}$. The spin Hall angle is assumed to be $\theta_{SH}=0.11$, also according to experiments by Garello et al. \cite{Garello_13,Garello_14}. Note that this value is also in the middle of the experimental bounds $0.056 \leq \theta_{SH} \leq 0.16$ estimated by Liu et al.\cite{Liu_11} and Garello et al.\cite{Garello_13}, and very close to the one deduced in\cite{Ryu_14}. A DMI parameter of $D=1.4\mathrm{mJ/m^{2}}$ is assumed, which is similar to the one experimentally deduced by Emori et al.\cite{Emori_13b}. The Gilbert damping is $\alpha=0.3$ as measured in \cite{Schellekens_13}. Several tests were also performed by varying these inputs within the range available in the experimental literature (see Supplementary Information).

\textbf{Acknowledgments}
This work was supported by project WALL, FP7-PEOPLE-2013-ITN 608031 from European Commission, project MAT2011-28532-C03-01 from Spanish government and projects SA163A12 and SA282U14 from Junta de Castilla y Leon.

\textbf{Author contributions}
E.M. and L.T. conceived and coordinated the project. N.P., L.T., M.H., V.R., S.M. and E.M. performed the micromagnetic simulations. E.M., L.T. and N.P. analyzed and interpreted the results. E.M. wrote the manuscript. All authors commented on the manuscript.           

\textbf{Additional information}
Supplementary Information is available. Correspondence and requests for materials should be addressed to E.M.  

\textbf{Competing financial interests}
The authors declare no competing financial interests. 

\end{document}